\documentclass[12pt]{article}
\usepackage[left=2cm, right=2cm, bottom=2cm ,top = 2cm, footskip=1cm]{geometry}
\usepackage[font=small,labelfont=bf]{caption}
\usepackage{subfig}

\usepackage{setspace}
\usepackage{lineno}

\let\counterwithin\relax
\usepackage{chngcntr}

\usepackage[hidelinks]{hyperref}
\usepackage[style=apa,maxbibnames=20,maxcitenames=2,backend=biber]{biblatex}
\usepackage{graphicx}
\usepackage{amsmath}

\usepackage[section]{placeins}

\usepackage{titling} % to add subtitle for one sentence summary
\newcommand{\subtitle}[1]{%
  \posttitle{%
    \par\end{center}    \begin{center}\normalsize#1\end{center}
    \vskip0.5em}%
}

\title{Brain Morphology Normative modelling platform for abnormality and Centile estimation: \textit{Brain MoNoCle}}

\author{Bethany Little$^{1,2}$, Nida Alyas$^1$, Alexander Surtees$^3$,\\
Gavin P Winston$^{4,5}$, John S Duncan$^4$, 
David A Cousins$^{2,6}$, John-Paul Taylor$^2$, \\
Peter Taylor$^{1,2,4}$, Karoline Leiberg$^1\ddagger$, Yujiang Wang$^{1,2,4*\ddagger}$}

\begin{document}

\maketitle

% author affiliations
\begin{enumerate}
\item{CNNP Lab (www.cnnp-lab.com), School of Computing, Newcastle University; Newcastle upon Tyne, United Kingdom.}
\item{Faculty of Medical Sciences, Newcastle University; Newcastle upon Tyne, United Kingdom.}
\item{Research Software Engineers, Newcastle University; Newcastle-upon-Tyne, United Kingdom.}
\item{UCL Queen Square Institute of Neurology; Queen Square, London, United Kingdom.}
\item{Department of Medicine (Division of Neurology), Queen’s University; Kingston, Canada.}
\item{Cumbria, Northumberland Tyne and Wear NHS Foundation Trust; Newcastle upon Tyne, United Kingdom.}
\end{enumerate}

\begin{center}
* Yujiang.Wang@newcastle.ac.uk\hspace{1.5cm}  
$\ddagger$ Joint senior authors
\end{center}

\newpage
\section*{Abstract}

%The preferred length of Research Article abstracts is 125 words or fewer, but a 250-word maximum is allowed for submission.
Normative models of brain structure estimate the effects of covariates such as age and sex using large samples of healthy controls. These models can then be applied to e.g. smaller clinical cohorts to distinguish disease effects from other covariates. However, these advanced statistical modelling approaches can be difficult to access, and processing large healthy cohorts is computationally demanding. Thus, accessible platforms with pre-trained normative models are needed.

We present such a platform for brain morphology analysis as an open-source web application \url{https://cnnplab.shinyapps.io/BrainMoNoCle/}, with six key features: (i) user-friendly web interface, (ii) individual and group outputs, (iii) multi-site analysis, (iv) regional and whole-brain analysis, (v) integration with existing tools, and (vi) featuring multiple morphology metrics.

Using a diverse sample of 3,276 healthy controls across 21 sites, we pre-trained normative models on various metrics. We validated the models with a small sample of individuals with bipolar disorder, showing outputs that aligned closely with existing literature only after applying our normative modelling. Using a cohort of people with temporal lobe epilepsy, we showed that individual-level abnormalities were in line with seizure lateralisation. Finally, with the ability to investigate multiple morphology measures in the same framework, we found that biological covariates are better explained in specific morphology measures, and for applications, only some measures are sensitive to the disease process.

Our platform offers a comprehensive framework to analyse brain morphology in clinical and research settings. Validations confirm the superiority of normative models and the advantage of investigating a range of brain morphology metrics together.

\section*{Keywords}

Normative model, brain structure, structural abnormality, morphology, bipolar disorder, temporal lobe epilepsy.

\newpage
%\section*{Main text:}
\section{Introduction}

Brain morphology, the study of the shape and size of brain structures, can be used to track healthy brain development and detect abnormalities associated with underlying disease processes. Normative modelling of brain morphology uses large and diverse datasets to estimate healthy variance across the lifespan. In neuroscience research, normative models can reliably remove biological and technical covariates from unseen data without removing e.g. disease effects \parencite{Pomponio2020}, which is especially valuable for small samples with a limited number of control subjects. Further, the ability of normative modelling to estimate abnormalities in individuals is crucial for clinical applications, as it enables systematic biomarker discovery and supports translational uses in diagnosis, stratification, and localisation \parencite{Loreto2024,Taylor2022, Little2023}. Therefore, normative modelling of brain morphology is an indispensable framework that should be available and accessible to all researchers.

To enable researchers without high levels of technical/statistical know-how, or resources for time and computationally demanding tasks, to benefit from the power of the normative modelling framework, a freely-available, pre-trained modelling platform is needed. The normative models should be based on large, diverse, healthy population data, and be easily-applied to new data. Recent efforts in this field include several open-source tools, some allowing users to upload new, unseen brain morphology data to a web interface and generate individual abnormality scores (e.g. z-scores or centiles) \parencite{Rutherford2023, Bethlehem2021, Ge2024}. Each of these tools have some advantages: for example, (i) being accessible as an online tool that can be easily-used without any need for software download/installation or writing/running scripts, (ii) providing individual and group-level outputs, (iii) multi-site data - as often seen in neuroscience research - can be analysed, (iv) analysing brain shape on whole hemispheres and smaller regions, (v) seamless integration with existing neuroimaging software such as FreeSurfer, and (vi) the option to explore a variety of metrics.

We present a normative modelling tool of brain morphology that combines all six key features in one open and web-based application: Brain MoNoCle (Brain Morphology Normative modelling platform for abnormality and Centile estimation). We included a large and diverse sample of 3,276 healthy controls across 21 sites to pre-train normative models in a variety of brain morphology measures that comprehensively quantify cortical shape, including three novel metrics that were only recently proposed \parencite{Wang2021}. As a first validation, we demonstrate how normative modelling improves reliability and reproducibility in a small clinical dataset of individuals with bipolar disorder (BD). We further validate our outputs in a dataset of temporal lobe epilepsy (TLE) at the group-level, and illustrate how individual patient abnormality scores agree with their seizure lateralisation. Finally, with the option to explore a variety of morphology metrics on our platform, we highlight the importance of investigating multiple metrics at the same time both for normative modelling itself, but also for clinical applications.

\section{Materials and methods}

\subsection{Normative data}

We collated 3T T1-weighted MRI scans from 3,276 healthy individuals from several large public and in-house datasets, detailed in Supplementary~\ref{norm_data_table} \parencite{Nigg2023, Himmelberg2023, Zareba2022, Nugent2022, VanEssen2013, Nooner2012, Shafto2014, Taylor2017, LaMontagne2019, Greene2018}. The age in the total dataset ranged from 5 to 95 years old; the age range and sex distribution for each study is illustrated in Figure \ref{norm_data}. Scanning protocols differed across, and sometimes within, datasets, which we corrected statistically in a later step. All studies had ethical approval from relevant institutional ethics boards and included written consent from participants. We present here the initial dataset included in v1.0 of our app; however, we aim to continuously add to our normative reference dataset. Users of our web platform should therefore check the latest summary of the dataset shown in the app when reporting their~own~results.

\subsection{Pre-processing}

T1-weighted MRI scans were pre-processed in FreeSurfer \parencite{Fischl2012} using the standard \textit{recon-all} pipeline, which includes removal of non-brain tissue, segmentation of grey and white matter surfaces, and cortical parcellation. We also ran the localGI pipeline \parencite{Schaer2012} to yield smoothed outer pial surfaces. The \textit{aparcstats2table} command was used to generate measures of cortical thickness, cortical volume, and pial surface area, for 68 brain regions according to the Desikan-Killiany parcellation atlas  \parencite{Desikan2006}. The version of FreeSurfer varied across datasets (see Supplementary~\ref{norm_data_table}), which was corrected during site/batch harmonisation.

%\begin{figure}
%	\centering
%	\includegraphics[width=\textwidth]{missing_fig.png}
%	\caption{\textbf{Normative model pipeline.}  \textbf{A}~T1-weighted MRI scans from healthy individuals were pre-processed using the standard recon-all pipeline in FreeSurfer. The aparcstats2table command was used to get measures of cortical thickness (CT), cortical volume (CV) and surface area (SA) for each brain region of interest (ROI). We used our cortical folding toolbox to calculate K, I, and S for each hemisphere. \textbf{B}~Generalised additive models for location, scale, and shape (GAMLSS) were then used to model healthy variance in each metric (for each brain region), accounting for non-linear age, sex, and site effects. \textbf{C}~The normative models are then used to predict abnormality scores in unseen datasets (in our example, people with mesial temporal lobe epilepsy (mTLE, n=) and healthy controls (HC; n=100). Here, healthy controls in the new dataset are used to correct for the new scanning site, before applying this correction to patient groups. Centiles and z-scores are then calculated.
%	}
%	\label{method}
%\end{figure}

\subsection{Cortical morphology measures}

The traditional morphological measures of cortical thickness, pial surface area, and exposed surface area are known to covary \parencite{Wang2016}. Failure to account for this covariance can lose and confuse information about the complex, folded shape of the brain. A recently developed framework proposed a universal scaling law of cortical folding that accounts for covariance between cortical thickness, pial surface area, and exposed surface area \parencite{Mota2015}. From this scaling law, three biologically interpretable independent components, K, I, and S can be derived for hemispheres, lobes, and individual regions \parencite{Wang2016, Wang2019, Wang2021, Leiberg2021}. The dimensionless measure K reflects tension acting on the cortex and is relatively preserved across species, but appears to be sensitive to ageing and disease processes \parencite{Wang2019, Wang2021, Leiberg2021}. Isometric term I is orthogonal and statistically independent to K and captures information about isometric size. S is a cross-product of K and I that captures all remaining information about shape, reflecting complexity of cortical folding. For example, if a cortical structure is isometrically rescaled in all dimensions, it changes I, but not K or S. K, I, and S are orthogonal and statistically independent to each other.

As an example, in TLE, these components captured structural changes that were not detected with traditional metrics \parencite{Wang2021}. K, I, and S therefore offer a novel re-conceptualisation of brain morphology measures that can detect nuanced morphological abnormalities.  We used the toolbox developed by \textcite{Wang2021} (\url{https://github.com/cnnp-lab/CorticalFoldingAnalysisTools}) to calculate K, I, and S for each hemisphere. 

\subsection{Quality control}

Some of the public datasets included quality control steps as part of the study design, which are reported in the original study publications. We detected outliers in the entire dataset statistically: we ran our \textit{gamlss} model described below for each structural metric for each region, and flagged outliers defined by residuals more than five median absolute deviations. In addition, we also detected outliers based on visual inspections of plots: for each dataset, we plotted each brain metric at the hemisphere level against age to flag outliers within each dataset. These were then cross-checked with outliers that were detected statistically. We excluded participants who were flagged as an outlier in any of these models; we performed listwise deletion rather than pairwise deletion so that the same normative reference dataset was used for each normative model, allowing comparisons of model statistics across models. 

\subsection{Exemplar clinical datasets}

To demonstrate the utility of our normative models in predicting abnormalities in patient groups and individuals, we included two exemplar clinical datasets. A sample of 133 adults with mesial TLE (mTLE; n=74 right hemisphere seizure onset; n=59 left hemisphere seizure onset) and 99 healthy controls (HC-mTLE) were acquired from the recent IDEAS dataset release \parencite{Taylor2024}, which was approved by a Research Ethics Committee (22/SC/0016). We also included a sample of 56 adults with bipolar disorder (BD) and 26 healthy controls (HC-BD) from the Bipolar Lithium Imaging and Spectroscopy Study (BLISS)  \parencite{Little2023, Smith2018}. The study was granted a favourable ethical opinion by a United Kingdom National Research Ethics Committee (14/NE/1135), and all participants provided written informed consent.  

Descriptive statistics for these datasets are summarised in Table \ref{demo_clinical}. See the respective publications for full details of each sample and neuroimaging pre-processing steps. 

\begin{table}[h!]
\centering
\begin{tabular}{c c c c c c}
 & \textbf{BD} & \textbf{HC-BD} & \textbf{mTLE left} & \textbf{mTLE right} & \textbf{HC-mTLE } \\ [2ex]
 \hline \\  [0.02ex]
 \textbf{n} & 56 & 26 & 74 & 59 & 99 \\ [2ex]
 \hline \\  [0.02ex]
 \textbf{Age [mean (SD)]} & 45.36 (12.15) & 48.46 (11.80) & 36.0 (11.2) & 38.2 (10.8) & 39.1 (12.1) \\ [2ex]
 \hline \\  [0.02ex]
 \textbf{Sex [n female (\%)]} & 35 (62\%) & 12 (46\%) & 43 (58\%) & 39 (66\%) & 62 (63\%) \\ [2ex]
 \hline
\end{tabular}
\caption{\textbf{Demographics of two clinical datasets.} SD=standard deviation; HC=healthy controls; BD=bipolar disorder; mTLE=mesial temporal lobe epilepsy.}
\label{demo_clinical}
\end{table}

\subsection{Software structure}

We designed our software as an R Shiny App, and it includes three aspects: (i) Pre-training normative models for each morphological measure, that can be readily used without re-fitting statistical models to normative data. (ii) At the back-end of the R Shiny App, computing z-scores and centiles for new unseen data based on the pre-trained models. (iii) At the front-end of the R Shiny App, providing the user with an intuitive interface that accepts outputs directly from existing neuroimaging software, such as FreeSurfer. Each aspect is summarised below, and more technical details can be found in the Supplementary.

\subsubsection{Pre-training normative models}

All brain metrics from the normative data were log-transformed before being used to train the normative models, so different metrics measuring different dimensionalities (e.g. thickness \textit{vs.} surface area) can be treated in the same way for the normative model. Generalized additive models for location scale and shape (GAMLSS) using the gamlss package (https://cran.r-project.org/web/packages/gamlss/) were used to simultaneously model the parameters (mean, standard deviation, skew, and kurtosis) of the distribution as response variables of the explanatory variables sex, age, and scanning site/batch. Specifically, in our model: the mean depends on sex (fixed effect), site (random effect), and a smooth function of age; the standard deviation depends on sex (fixed effect), site (random effect), and a smooth function of age; the skew depends on sex (fixed effect) and a smooth function of age; and the kurtosis depends on a smooth function of age. See the Supplementary~\ref{supplStatModel} for a more detailed description and justification of the statistical model. Normative models were fitted independently for each region (from the Desikan-Killiany atlas) and hemisphere, and each morphometric measure. Residuals were retained for later visualisations.

\subsubsection{Predicting abnormalities in unseen data}

The pre-trained normative models are implemented on the R Shiny App back-end to  score new, unseen individuals. A healthy control (HC) cohort from the new unseen site/batch is currently required.

First, we predicted the distribution parameters based on the HCs in each new site/batch and calculated residuals relative to one of the normative scanning sites. The mean of the residuals from the HCs in the unseen data was then used to calculate the site-specific offset needed to harmonise the unseen dataset with the normative data. 

To obtain z-score, we then calculated the residuals for each individual in the new unseen data relative to their site mean and divided by a standard deviation. The latter is calculated as follows: if there are less than 30 HCs, we used the average standard deviation seen across normative data sites/batches in the pre-trained normative models. We estimated the standard deviation from the unseen HCs only if there were 30 or more HCs in the dataset. This approach ensures accurate estimations of new sites' standard deviations, as a sample size of 30 provides a 55\% probability of being within 10\% accuracy and a 95\% probability of being within 25\% accuracy~\parencite{Schillaci2022}.

The site-specific mean, (site-specific) standard deviation, skew, and kurtosis from the pre-trained normative model, were used to calculate centiles. See the Supplementary~\ref{supplStatModel} for a detailed description of the statistical pipeline.  

\subsubsection{Using the Brain MoNoCle web user interface}
To run our pipeline to predict abnormalities in unseen data as described above, we used our web platform Brain MoNoCle. Users can follow the same steps to run the pipeline on their own data. 

First, we uploaded pre-processed brain imaging data tables. For traditional brain imaging metrics, data should be pre-processed using FreeSurfer (e.g., the standard \textit{recon-all} command) and the structural metrics for each hemisphere should be exported as csv files using the \textit{aparcstats2table} command; then the csv file for each hemisphere can be directly uploaded to our web interface. For morphology measures of independent components, the data tables from our cortical folding toolbox (\url{https://github.com/cnnp-lab/CorticalFoldingAnalysisTools}) can be directly uploaded. We also uploaded meta-data in a csv file containing subject IDs, age, sex, group, dataset, scanning site, and session.  After selecting 'Run Model' to start the analysis, z-scores, group summary statistics, and centiles are available to view and download as csv files, using the tabs in the main panel. Users can export plots by selecting the `Brain plot' and `Scatter plot' tabs. A html report is available to download using the `Report' tab. 

\subsection{Statistical analysis}

All statistical analysis was performed in R Studio v4.3.2. Each test and associated sample size is stated in the results section. 

%\subsection{Case-control comparison for bipolar disorder}
%To compare the outputs of our normative modelling pipeline with a traditional case-control pipeline, we also ran a traditional stream of analyses on cortical thickness differences in BD and their matched HC group. We extracted cortical thickness (CT) for 68 ROIs using FreeSurfer \textit{aparcstats2table} command and log transformed CT data. Age and sex were regressed out in the control group using robust regression for each ROI; age and sex were then regressed out of BD data using the regression model that was estimated using the HCs. Residuals following age and sex regression were then standardised (z-scored) based on the HC mean and SD for each ROI. We then tested the magnitude of group differences using Cohen’s d.

\section{Results}

\subsection{Pre-trained normative models on web platform}

\begin{figure}[h]
	\centering
	\includegraphics[scale=1]{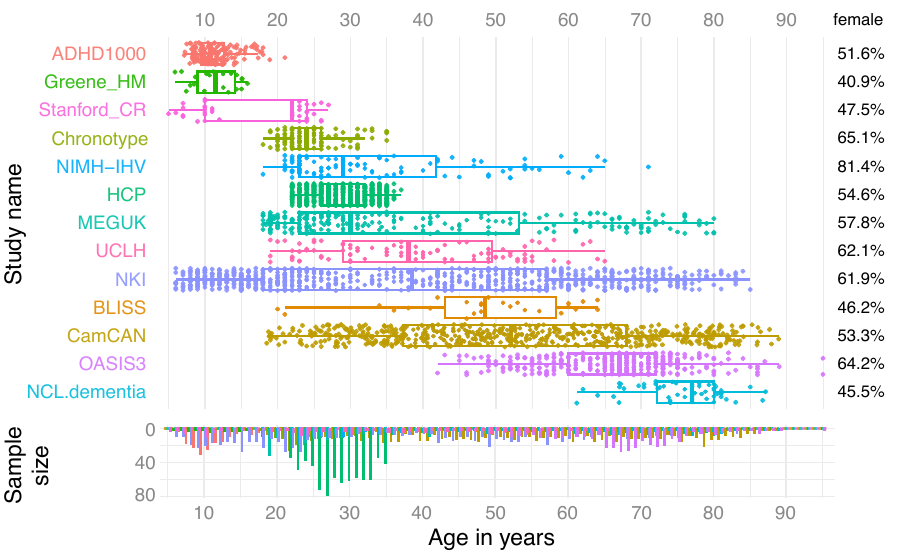}
	\caption{\textbf{Demographics of the data underlying the normative model.} Age distributions and proportion of female participants are shown for each study.
	}
	\label{norm_data}
\end{figure}

To pre-train normative models, we used data from 3,276 healthy individuals from several large public and in-house datasets, detailed in Supplementary~\ref{norm_data_table} \parencite{Nigg2023, Himmelberg2023, Zareba2022, Nugent2022, VanEssen2013, Nooner2012, Shafto2014, Taylor2017, LaMontagne2019, Greene2018}. We focused on including a variety of scanning protocols/sites (total 21 sites) to enable mixed-effect modelling, and we achieved a larger overall sample size than recommended by previous normative models \parencite{Ge2024}. We performed a subsampling analysis that showed our models stabilised, see Supplementary~\ref{supplStable} for details. The age in the total normative dataset ranged from 5 to 95 years old and is shown by data source in~Fig.~\ref{norm_data}. With this data we pre-trained our normative models for the whole hemisphere, each brain region, and morphology metric (see Methods for details of statistical models). We incorporated these pre-trained normative models on a web platform (Brain MoNoCle) to allow users to upload their own datasets to find morphological abnormalities in individuals and groups. In the following, we will validate our normative modelling framework and web platform outputs in a variety of ways, and demonstrate some biological insight.

\subsection{Cortical thickness abnormalities in a sample of patients with bipolar disorder more closely match previous findings when using the normative model}

\begin{figure}
	\centering
	\includegraphics[scale=0.85]{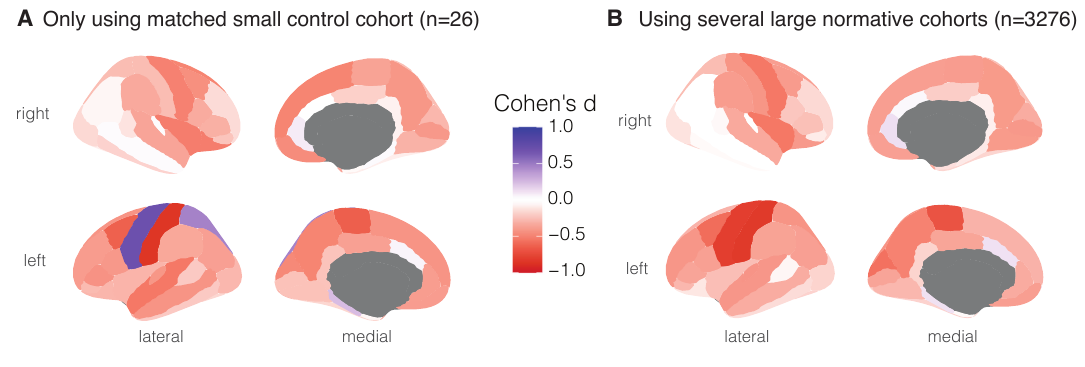}
	\caption{\textbf{Alterations in cortical thickness associated with bipolar disorder derived from a case-control study \textit{vs.} normative modelling.}  Group-level abnormalities in cortical thickness in n=56 people with bipolar disorder for: \textbf{A}~a small, matched control group (n=26); and \textbf{B}~the normative reference population (n=3,276).
	}
	\label{BP_output}
\end{figure}

To validate the normative modelling framework and outputs, we first investigated the group-level differences in a small, well-defined clinical cohort of bipolar disorder (BD, n=56) and matched controls (BD-HC, n=26). We specifically wanted to see the difference in outputs between using a traditional case-control comparison approach only using the matched controls (Figure~\ref{BP_output}~A) \textit{vs.} using our normative model instead (Figure~\ref{BP_output}~B). When using the small BD-HC group, effect sizes (Cohen's d) suggested that cortical thinning was greatest in the left post-central gyrus (d=-0.85) and that the cortex was thicker in BD in the left pre-central gyrus (d=0.70). However, when using the normative model pipeline, the same sample showed similar thinning in the left post-central gyrus (d=-0.83), but cortical \textit{thinning} in the left pre-central gyrus (d=-0.8). The latter of these findings, obtained through normative modelling, is more in line with previous findings from a large sample ENIGMA study showing cortical thinning across the cortex \parencite{Hibar2018}.

\subsection{Group-level cortical thickness abnormalities in mesial temporal lobe epilepsy agree with previous findings}

\begin{figure}[h!]
	\centering
	\includegraphics[scale=1]{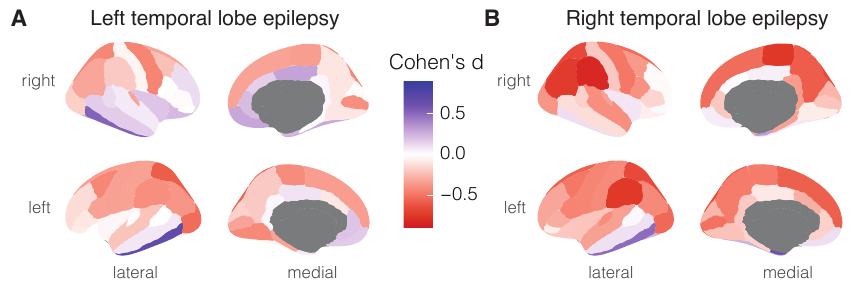}
	\caption{\textbf{Group-level output for mesial temporal lobe epilepsy cohort after normative modelling.} Group-level summary of abnormalities in cortical thickness for left mTLE (n=74, \textbf{A}) and right mTLE (n=59, \textbf{B}), showing Cohen's d effect size for each cortical region. }
	\label{tle_outputs}
\end{figure}

We validated our normative model outputs in a large sample of individuals with mTLE (74 left mTLE, 59 right mTLE) and matched controls (n=99) \parencite{Taylor2024}. Figure~\ref{tle_outputs} shows cortical thickness abnormality estimates for right mTLE and left mTLE groups. We found widespread cortical thinning, especially in the right mTLE group, in cortical regions including the precentral gyrus, supramarginal gyrus, and inferior parietal gyrus. This result reproduces both the findings from the IDEAS and ENIGMA-epilepsy studies. We quantitatively assessed this by correlating the effect sizes for each brain region generated by Brain MoNoCle in our mTLE sample with the effect sizes reported in the ENIGMA study \parencite{Whelan2018}; results showed agreement between both sets of effect sizes (see supplementary materials~\ref{supplENIGMAcorr}) . %Average abnormalities across subjects in each structural metric computed for cortical hemispheres can be found in the supplementary material.

As a supplementary step, we compared the z-scores for each brain region produced by Brain MoNoCle with z-scores produced by a similar normative modelling platform, CentileBrain, using the IDEAS healthy control group (n=99). Results show good agreement between both apps (correlation larger than 0.75 in approx. 80\% of brain regions, see supplementary materials~\ref{supplCentBrain}).

\subsection{Individual-level abnormalities in certain measures agree with clinical lateralisation of seizure onset}

\begin{figure}[h!]
	\centering
	\includegraphics[scale=1]{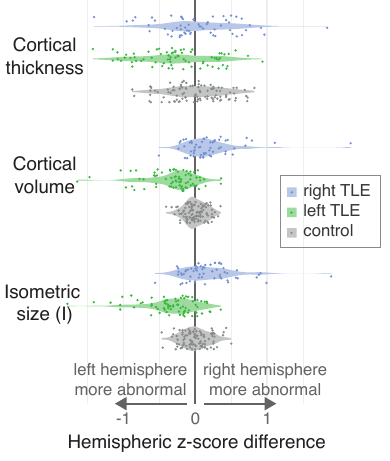}
	\caption{\textbf{Individual-level z-scores after normative modelling for mTLE cohort.} Difference in hemisphere-level z-score between left and right hemisphere is shown in controls and right/left TLE subgroups for three example morphological measures. Individual subjects are shown as single data points, distributions of subjects are displayed as violin plots.}
	\label{lateralisation}
\end{figure}

To validate individual-level outputs and abnormalities, we used seizure lateralisation from the IDEAS dataset \parencite{Taylor2024}. For each subject, we extracted the z-score difference between left and right hemisphere in cortical thickness and other metrics (Figure~\ref{lateralisation}). Controls, as expected, had a distribution around zero in all measures after regressing out healthy biological covariates.

%with a standard deviation of 0.38 for thickness. 0.16 for volume and pial surface area, 0.19 for I, . 

The measure most frequently-used in cortical morphometry, cortical thickness, did not lateralise individual patients  well, and most patient z-scores differences were within the same range and distribution as the controls. 

Given that our normative modelling platform offers the ability to analyse multiple morphology measures, and given that cortical thickness is known to covary with other measures (such as surface area and volume), we investigated all metrics implemented on the platform. This included three statistically independent novel measures. We demonstrated that both cortical volume and I - our novel morphometric for isometric size - were best at lateralising at the hemisphere level (Figure~\ref{lateralisation}). Specifically, for cortical volume, most (78.2\%) patients had a z-score difference greater/smaller than zero indicating lateralisation in agreement with clinical metadata. For I, 75.9\% of patients showed the correct laterality. Further, 36.1\% of patients were outside of 2 standard deviations of the control for cortical volume, and 32.3\% for I. Confusion matrices showing predictive performance for lateralisation using the sign of the z-score difference between left and right hemisphere can be found in the Supplementary~\ref{supplLat}. Overall performance accuracy was larger than 0.75 for both volume and I.

% In figure \ref{tle_outputs} C), we show an web platform output of an individual map of regional thickness/volume abnormalities one example individual with left lateralisation (Figure \ref{tle_outputs} C).

\subsection{Covariates explain more variance in independent component K than in other structural MRI measures}

Given the observed specificity in particular measures for seizure lateralisation, we explored the differences between morphological measures further to establish a baseline for future applications. To this end, we investigated the normative models accounting for age, sex and scanning site for each measure. 

\begin{figure}[h]
	\centering
	\includegraphics[scale=0.9]{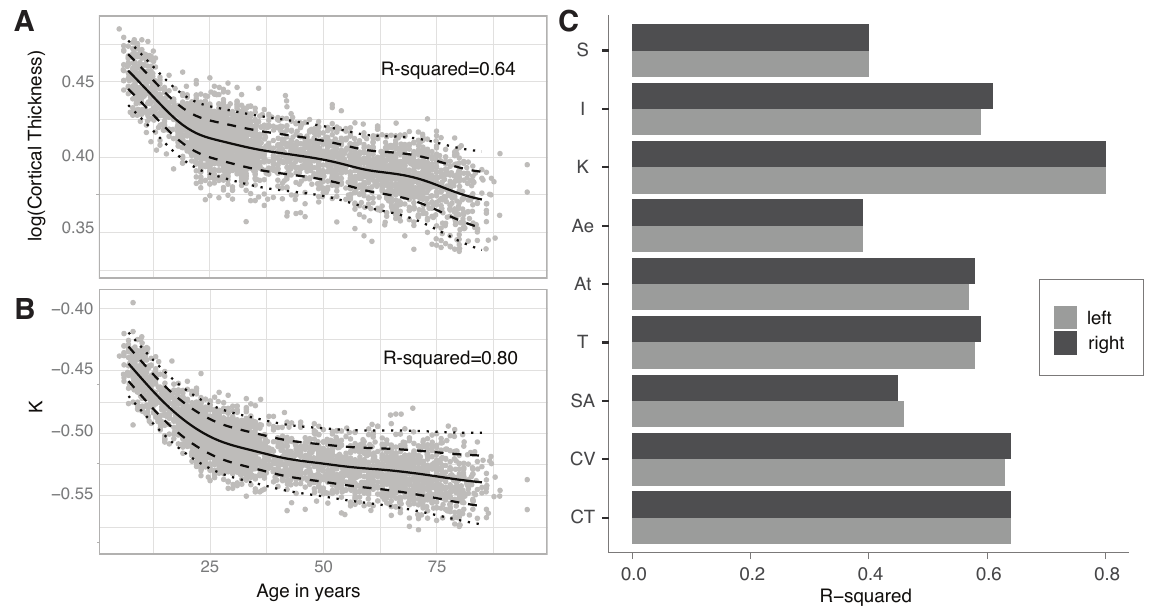}
	\caption{\textbf{Variance explained by normative model in each morphometric.} \textbf{A~\&~B}~Harmonised normative data (grey dots) and predicted model centiles of mean cortical thickness and K across the lifespan (n=3,276).
 \textbf{C}~Model fit statistics $R^2$ for each metric and hemisphere. CT, CV, and SA are structural metrics estimated using FreeSurfer; T, At, Ae, K, I and S are structural metrics estimated using the Cortical Folding toolbox. CT=cortical thickness, CV=cortical volume, SA=surface area; T=average thickness, At=total pial surface area, Ae=exposed surface area. 
	}
	\label{Normrsq}
\end{figure}

Figure~\ref{Normrsq}~A and B show the fitted normative model over age for two example measures: cortical thickness and K - a novel independent morphometric that is known to change with age \parencite{Wang2021}. Both thickness and K decrease over age with steeper declines in early and later life. 

To compare morphology measures more directly, we obtained the $R^2$ of the normative model fit for each measure Figure~\ref{Normrsq}~C. All measures used the same statistical model formulas of age, sex and site. Out of all the measures implemented, K shows the best model fit ($R^2=0.8$ for both left and right hemisphere), superior to all other metrics with $R^2$ around or below 0.6.

\section{Discussion}
\textit{Summary:} Brain MoNoCle is a user-friendly online normative modelling platform for brain morphology analysis. The platform combines and unifies the most frequently-requested and desired features of existing approaches and toolboxes in one, importantly including the option to analyse multiple morphology measures under one framework. We validated our normative models and platform outputs in clinical cohorts through a series of tests, including replicating previous findings from ENIGMA studies. We also provided an individual-level output validation in a sample of mTLE by demonstrating agreement of our outputs with the clinical seizure lateralisation. Particularly, we highlighted that both biological covariates, as well as disease processes, are uniquely expressed in different morphological measures. This implies that brain normative modelling should be performed in a range of measures to be useful for brain morphology analysis in health and disease.

\textit{Validations:} We demonstrated how our pipeline can be applied to clinical datasets and what outputs can be obtained in a sample of people with BD, and a sample of people with mTLE. Compared to the traditional case-control pipeline, in which BD patients were compared to matched healthy controls, our normative pipeline yielded abnormality estimates that were more in line with previous research, for example a large-scale ENIGMA study of cortical thickness in BD, despite a relatively small patient sample  \parencite{Hibar2018}. In mTLE, we showed results similar to a recent report of the data-release \parencite{Taylor2024}. In particular, we did not find large abnormalities in mTLE patients; this may be because we only show cortical data and mTLE is associated with structural changes in subcortical regions, such as the hippocampus \parencite{Whelan2018}. We also tested for lateralisation of the hemispheric abnormalities, and found a generally good agreement with clinical metadata, despite only using cortical data. The reported effect sizes compared to controls are in line with previous reports of lateralisation using cortical information only \parencite{PUSTINA201520, Whelan2018}. We conclude that our normative models provide reasonable outputs in small and large samples, and that individual-level outputs are also in line with expectations. We hope Brain MoNoCle will be helpful in future analysis of cortical morphology. 

\textit{Normative sample:} We trained our models using healthy control data from several large databases (Supplementary~\ref{norm_data_table} and \ref{supplaackn}), similar to other normative modelling approaches. Our model is stable with the current sample size of 3,276 subjects. However, in mixed effect modelling, the number of levels in the random effect (i.e., the number of sites) is also critical. It is recommended to use a fixed effect if there are fewer than 10 levels, and to apply caution with \~20 levels as a random effect. We included 21 sites, and suggest that future work should prioritize increasing the number of sites in normative datasets, with diverse global representation, rather than focusing solely on total subject count.

\textit{Methodological advance:} Our normative models and web platform is an addition to existing free toolboxes for modelling cortical morphometry \parencite{Rutherford2023, Bethlehem2021, Ge2024, Manjón2016, attye2024}. Brain MoNoCle, however, differs to all of these in some ways, including: No requirement for any coding or running scripts from the user; no need to download software; outputs include group-level analysis compared to controls as well as individual abnormalities as z-scores and centiles; analysis is on full hemispheres and regions; and outputs are visualised in plots and available as tables of z-scores and centiles. There are differences in the underlying models as well. For example, our model harmonises scanning site effects for mean and variance, avoiding separate steps as found in Combat \parencite{Johnson2007}, which makes separate assumptions. Specifically, Combat assumes that after removing covariate effects on the mean, the data is normally distributed, and site effects on mean and variance are estimated based on this assumption. However, this is often not the case, since data may follow a non-normal distribution with higher distribution parameters of skew and kurtosis also depending on covariates. Further, in our model information is pooled from the male and female populations to directly estimate sex covariate effects; and we use flexible smooth terms for age effects and explicitly model skew and kurtosis with GAMLSS. Lastly, our web platform includes normative models of a range of metrics, including traditional measures like thickness, volume, and pial surface area, but also independent morphometrics which account for the covariance of those measures.

\textit{New biological insight:} Through our exploration of multiple cortical morphometrics, we were able to compare normative models for traditional measures, such as cortical thickness and surface area, but also novel statistically independent morphometrics. We found that one of these novel morphometrics ``K'' (also termed ``tension component'') displayed a far superior performance as a normative model of age and sex with $R^2=0.8$ compared to other morphometrics that achieve $R^2$ between 0.4 and 0.6. This observation has two implications: firstly, there might be better morphometrics to use to model age and sex effects, and extract disease-specific effects, as alluded to in the first paper proposing ``K'' as a novel morphometric \parencite{Wang2021}. Secondly, traditional morphometrics clearly have residual unexplained variance due to their statistical interdependence. This implies that investigations of measures such as cortical thickness, and surface area should consider their covariance, rather than interpreting them in isolation. The cortex, as a biological structure obeying physical constraints, clearly does not have independent processes to develop its thickness vs. surface area vs. overall size. Our platform, offering analysis streams for all traditional morphometrics and novel morphometrics therefore serves as a starting point for future statistically robust analyses of brain morphology.

\textit{Roadmap for future development:} We have three concrete developments planned for our normative modelling platform. First, we will incorporate recently-proposed multiscale morphometrics \parencite{Chen2022, Wang2022,Leiberg2023} to allow users to access the most recent cutting-edge developments in morphological analysis. Second, we currently use one Freesurfer parcellation of the brain to analyse finer regions. We plan to incorporate more atlases, and in the same step, incorporate the possibility to jointly model related regions (e.g. neighbouring regions) to increase robustness of the model. We also note that vertex-wise data and surface-based statistics would be a useful addition. Third, with the increasing availability of longitudinal data, we plan to extend our normative model to accept multi-session longitudinal clinical datasets and statistically account for these adequately (see e.g. \textcite{buvckova2023} for some suggestions). Further, we will add more normative data from diverse geographical areas. We will also implement analysis capacity to compare morphology measures more directly on the web platform. Finally, we aim to add more structural metrics such as subcortical volumes, and integrate output from other neuroimaging tools, such as CIVET and volBrain \parencite{Manjón2016, lee2006}.

%\section{Contributors}

%\section{Declaration of interests}

\newpage

\section{Data and materials availability}
Normative data may be available at the discretion of the data holders, please see the website of individual datasets for more information. The subset of the IDEAS mesial TLE dataset is freely available with the associated paper \parencite{Taylor2024}.

\section{Author contributions} 

Conceptualization: BL, KL, and YW. 
Methodology: BL, KL, and YW. 
Investigation: BL, NA, AS, KL, and YW. 
Visualization: BL, KL, PT, and YW. 
Funding acquisition: GPW, JSD, DAC, PT, and YW. 
Project administration: BL and YW. 
Supervision: YW. 
Writing – original draft: BL. 
Writing – review \& editing: BL, NA, AS, GPW, JSD, JPT, DAC, PT, KL, and YW.

\section{Funding}

BL, YW, and KL were supported by the EPSRC (EP/Y016009/1). YW and PT were supported by UKRI Future Leaders Fellowships (MR/V026569/1, MR/T04294X/1). The Bipolar Lithium Imaging and Spectroscopy Study (BLISS) project was funded by the Medical Research Council (Clinician Scientist Fellowship BH135495 to DAC). The normative MEG UK data collection was supported by an MRC UK MEG Partnership Grant, MR/K005464/1. GPW and the collection of control data for the UCLH dataset were supported by the MRC (G0802012 and MR/M00841X/1) and the NIHR UCLH/UCL Biomedical Research Centre. JP-T and DC were supported by the Newcastle NIHR Biomedical Research Centre. 

The funders did not have a role in study conception, design, data collection, data analysis, interpretation of the data, preparation of the article, or article submission.

\section{Declaration of Competing interests}
There are no competing interests to disclose.

\section{Acknowledgements}

We thank members of the Computational Neurology, Neuroscience \& Psychiatry Lab (www.cnnplab.com) for discussions on the analysis and manuscript. The full list of acknowledgements for the normative data is provided in the supplementary materials. We thank all participants who generously contributed their time and data.

\newpage

%\bibliography{ref}
\printbibliography

%%%%%%%%%% %%%%%%%%%% SUPPLEMENTARY %%%%%%%%%% %%%%%%%%%% 

\newpage

\section*{Supplementary Materials}

\renewcommand{\thefigure}{S\arabic{figure}}
\setcounter{figure}{0}
\counterwithin{figure}{section}
\counterwithin{table}{section}
\renewcommand\thesection{S\arabic{section}}
\setcounter{section}{0}

\section{Normative data} \label{norm_data_table}

\begin{table}[h!]
\footnotesize
\centering
\begin{tabular}{c c c c c c}
 \textbf{Dataset} & \textbf{Sample size} & \textbf{Age in years} & \textbf{Sex} & \textbf{Location} & \textbf{FreeSurfer} \\ [2ex]
\textbf{(site)}  & \textbf{[n]} & \textbf{[median(range)]} & \textbf{[n female (\%)]} & \textbf{[Country]} & \textbf{[Version]} \\ [2ex]
 \hline \\  [0.02ex]
 ADHD1000 & 153 & 11 (7 - 21) & 79 (51.6\%) & USA & 7.3.2 \\ [2ex]
 %\hline \\  [0.02ex]
 BLISS & 26 & 49 (20 - 64) & 12 (46.2\%) & UK & 6.0.1 \\ [2ex]
 %\hline \\  [0.02ex]
 CamCAN (s1) & 495 & 50 (19 - 89) & 259 (52.3\%) & UK & 5.3.0 \\ [2ex]
 %\hline \\  [0.02ex]
 CamCAN (s2) & 83 & 59 (19 - 86) & 49 (59.0\%) & UK & 5.3.0 \\ [2ex]
 %\hline \\  [0.02ex]
 Chronotype & 126 & 24 (18 - 35) & 82 (65.1\%) & Poland & 7.3.2 \\ [2ex]
 %\hline \\  [0.02ex]
 Greene-HM & 22 & 11 (6 - 16) & 9 (40.9\%) & USA & 7.3.2 \\ [2ex]
 %\hline \\  [0.02ex]
 HCP & 678 & 29 (22 - 37) & 370 (54.6\%) & USA & 5.2 \\ [2ex]
 %\hline \\  [0.02ex]
 MEGUK (Aston1) & 29 & 23 (18 - 49) & 25 (86.2\%) & UK & 7.3.2 \\ [2ex]
 %\hline \\  [0.02ex]
 MEGUK (Aston2) & 69 & 33 (18 - 63) & 39 (56.5\%) & UK & 7.3.2 \\ [2ex]
 %\hline \\  [0.02ex]
 MEGUK (Cambridge) & 71 & 42 (19 - 80) & 37 (52.1\%) & UK & 7.3.2 \\ [2ex]
 %\hline \\  [0.02ex]
 MEGUK (Glasgow) & 24 & 26 (18 - 34) & 13 (54.2\%) & UK & 7.3.2 \\ [2ex]
 %\hline \\  [0.02ex]
 MEGUK (Oxford) & 63 & 40 (20 - 80) & 34 (54.0\%) & UK & 7.3.2 \\ [2ex]
 %\hline \\  [0.02ex]
 NCL-dementia (s1) & 29 & 77 (62 - 85) & 11 (37.9\%) & UK & 6.0.1 \\ [2ex]
 %\hline \\  [0.02ex]
 NCL-dementia (s2) & 26 & 74 (61 - 87) & 14 (53.8\%) & UK & 6.0.1 \\ [2ex]
 %\hline \\  [0.02ex]
 NIMH-IHV (s1) & 48 & 27 (18 - 63) & 35 (72.9\%) & USA & 7.3.2 \\ [2ex]
 %\hline \\  [0.02ex]
 NIMH-IHV (s2) & 54 & 33 (21 - 71) & 48 (88.9\%) & USA & 7.3.2 \\ [2ex]
 %\hline \\  [0.02ex]
 NKI & 790 & 39 (6 - 85) & 489 (61.9\%) & USA & 6.0.1 \\ [2ex]
 %\hline \\  [0.02ex]
 OASIS3 & 355 & 67 (42 - 95) & 228 (64.2\%) & USA & 7.3.2 \\ [2ex]
 %\hline \\  [0.02ex]
 Stanford-CR & 40 & 22 (5 - 27) & 19 (47.5\%) & USA & 7.3.2 \\ [2ex]
 %\hline \\  [0.02ex]
 UCLH (s1) & 28 & 37 (19 - 64) & 16 (57.1\%) & UK & 7.3.2 \\ [2ex]
 %\hline \\  [0.02ex]
 UCLH (s3) & 67 & 39 (19 - 65) & 43 (64.2\%) & UK & 7.3.2 \\ [2ex]
 %\hline \\  [0.02ex]
 \hline \\  [0.02ex]
 Total & 3,276 & 33 (5 - 95) & 1,911 (58.3\%) \\ [2ex]
 \hline \\  [0.02ex]
 \hline
\end{tabular}
\caption{\textbf{Descriptive statistics of each study dataset included in the normative model.}}
\label{NM_data}
\end{table}

\clearpage
\section{Statistical modelling} \label{supplStatModel}

\subsection{Model formula}

For the normative model, we assume the morphological data, measured in all metrics including cortical thickness, volume, surface area, tension $K$, shape $S$, and isometric size $I$, follow a flexible sinh-arcsinh (shash) distribution, which allows for non-normal modelling of the data in which the first four moments of the distribution can vary as functions of explanatory variables. All metrics are transformed to a logscale before statistical modelling. We use gamlss (\url{https://www.rdocumentation.org/packages/gamlss/versions/5.4-12/topics/gamlss}) to simultaneously model the parameters (first four moments) of the distribution as response variables of the explanatory variables sex, age, and scanning site. We use the model formulae:

\begin{equation*} \label{eq1}
\begin{split}
\mu & \sim 1 + sex + s(age) + (1 | site) \\
\sigma & \sim 1 + sex + s(age) + (1 | site) \\
\nu & \sim 1 + sex + s(age) \\
\tau & \sim 1 + s(age)
\end{split}
\end{equation*}

\begin{itemize}
    \item The mean ($\mu$) depends on sex (fixed effect), site (random effect), and a smooth function of age.
    \item The standard deviation ($\sigma$) depends on sex (fixed effect), site (random effect), and a smooth function of age.
    \item The skew ($\nu$) depends on sex (fixed effect) and a smooth function of age.
    \item The kurtosis ($\tau$) depends on a smooth function of age.
\end{itemize}

We fitted this model to the normative datasets ($n\sim3500$) for each metric and cortical region independently.

\begin{figure}
  \centering
  \includegraphics[width=\textwidth]{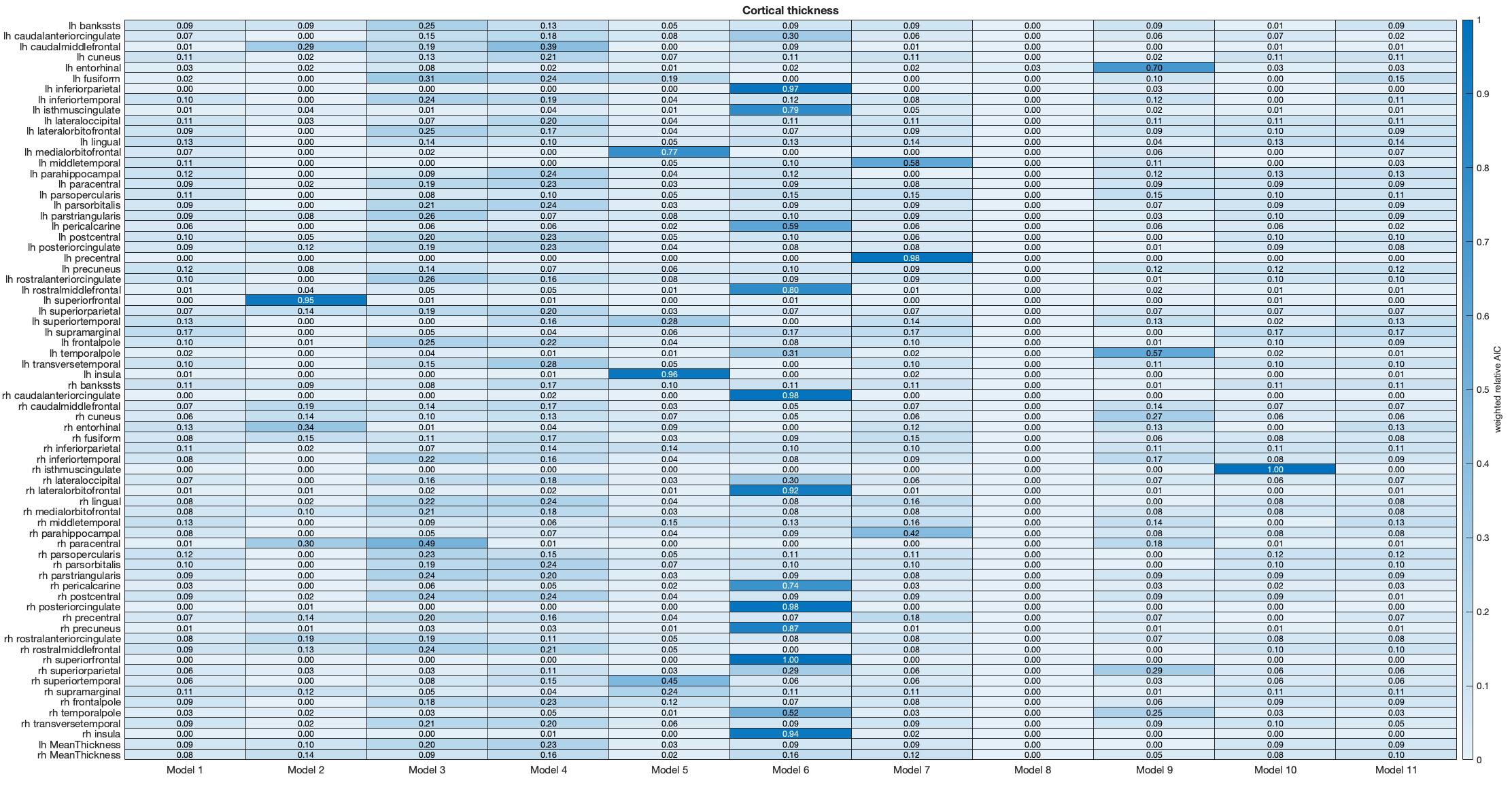}
%  \includepdf{wAIC_CT.png}
  \caption{\textbf{Weighted relative AIC of 11 gamlss models fitted for cortical thickness.} Darker colour indicates higher probability that the model is optimal for that metric and region out of the 11 models tested. Rows correspond to 68 regions (Desikan-Killiany atlas) as well as left and right hemisphere mean thickness. Columns correspond to 11 model formulae.}
  \label{CT}
\end{figure}

\begin{figure}
  \centering
  \includegraphics[width=\textwidth]{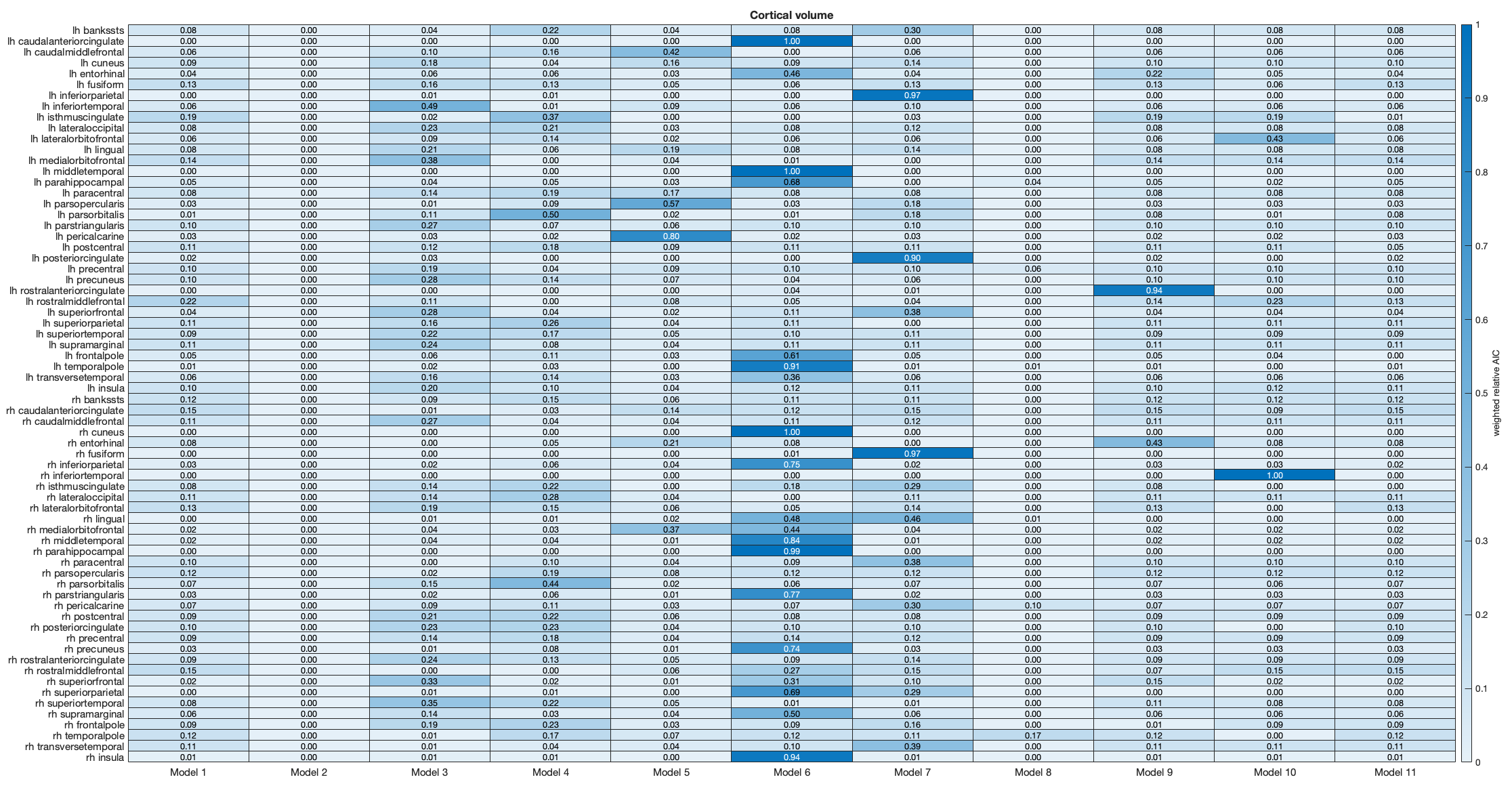}
%  \includepdf{wAIC_CT.png}
  \caption{\textbf{Weighted relative AIC of 11 gamlss models fitted for cortical volume.} Darker colour indicates higher probability that the model is optimal for that metric and region out of the 11 models tested. Rows correspond to 68 regions (Desikan-Killiany atlas). Columns correspond to 11 model formulae.}
  \label{CV}
\end{figure}

\begin{figure}
  \centering
  \includegraphics[width=\textwidth]{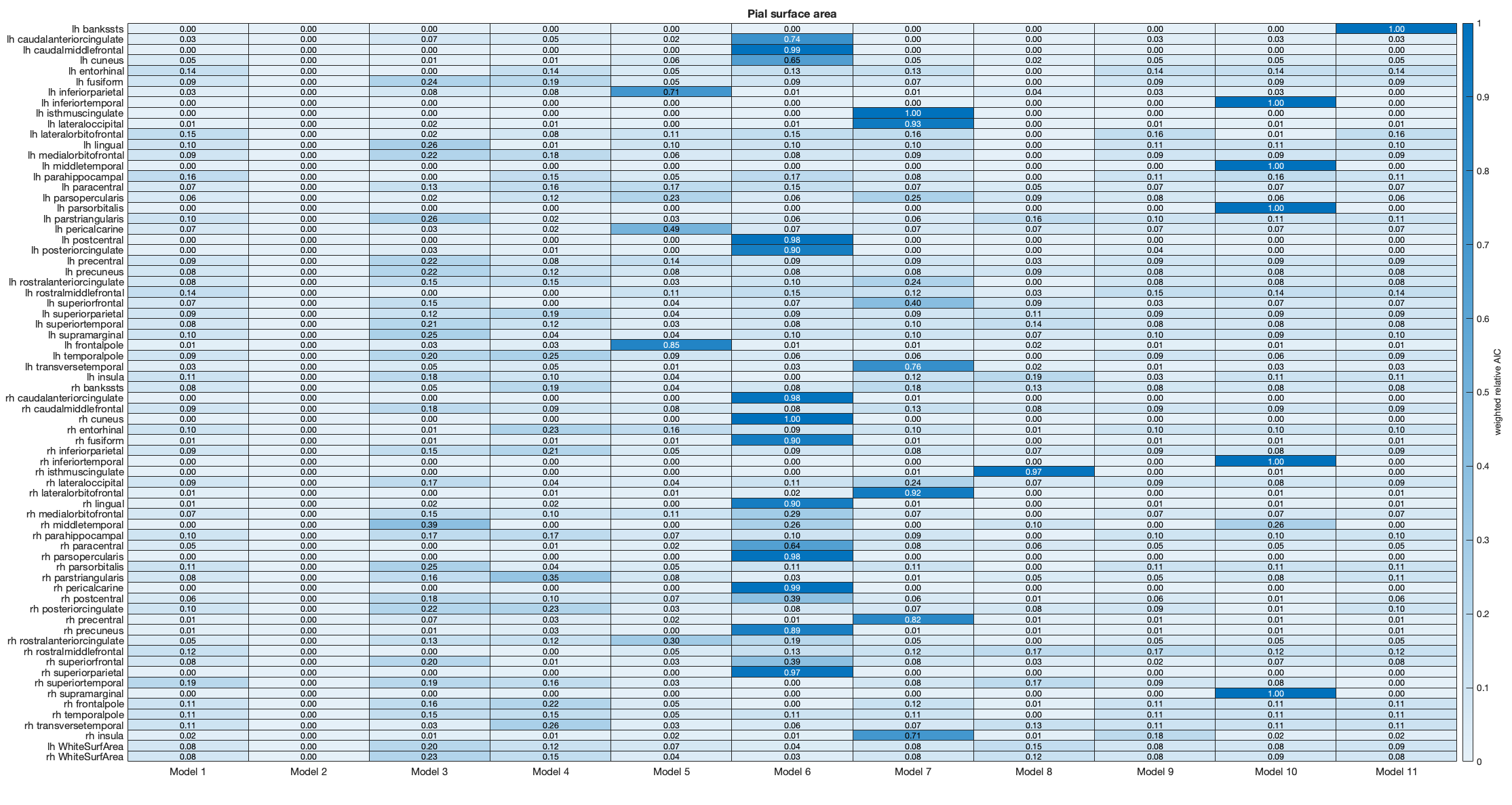}
%  \includepdf{wAIC_CT.png}
  \caption{\textbf{Weighted relative AIC of 11 gamlss models fitted for pial surface area.} Darker colour indicates higher probability that the model is optimal for that metric and region out of the 11 models tested. Rows correspond to 68 regions (Desikan-Killiany atlas) as well as left and right hemisphere total surface area. Columns correspond to 11 model formulae.}
  \label{SA}
\end{figure}

To test our model formula, we fitted alternative models to our data, changing one model term at a time. This resulted in the comparison of 11 models in total. The model as described above is considered the default (Model~1 in figures \ref{CT}, \ref{CV}, \ref{SA}). In models 2-11, the formulae of all parameters were kept the same as in model~1, except as described below:

\begin{equation*} \label{eq2}
\begin{split}
\text{Model 2: }\mu & \sim 1 + s(age) + (1 | site) \\
\text{Model 3: }\sigma & \sim 1 + s(age) + (1 | site) \\
\text{Model 4: }\nu & \sim 1 + s(age) \\
\text{Model 5: }\tau & \sim 1 + sex + s(age) \\
\text{Model 6: }\nu & \sim 1 + sex +  s(age) + (1 | site) \\
\text{Model 7: }\tau & \sim 1 + s(age) + (1 | site) \\
\text{Model 8: }\mu & \sim 1 + sex + age + (1 | site) \\
\text{Model 9: }\sigma & \sim 1 + sex + age + (1 | site) \\
\text{Model 10: }\nu & \sim 1 + sex + age \\
\text{Model 11: }\tau & \sim 1 + age
\end{split}
\end{equation*}

I.e., models 2-5 removed/introduced a sex term to the four parameters, models 6-7 introduced a site effect to the skew and kurtosis, and models 8-11 simplified the smooth function of age to a linear term in all four parameters.
We computed AICs (Akaike information criterion) for each model fit in each metric and cortical region, and assessed the optimal model for each metric and region by computing weighted relative AICs (figures \ref{CT}, \ref{CV}, \ref{SA}).

Generally, our model fitted the data better or similarly well as the alternative models. In a subset of cortical regions, model~6 with a random site effect on the skew had a significantly better fit. However, we opted not to include this term in our final model for two reasons: a), to keep the model formula consistent across metrics and regions for comparability, and b) to allow us to apply the normative model to new datasets with small healthy control samples, which would make an accurate estimation of site-specific skew impossible.

\newpage

\subsection{Application to new data}

\begin{figure}[h!]
\centering
  \includegraphics[scale=1]{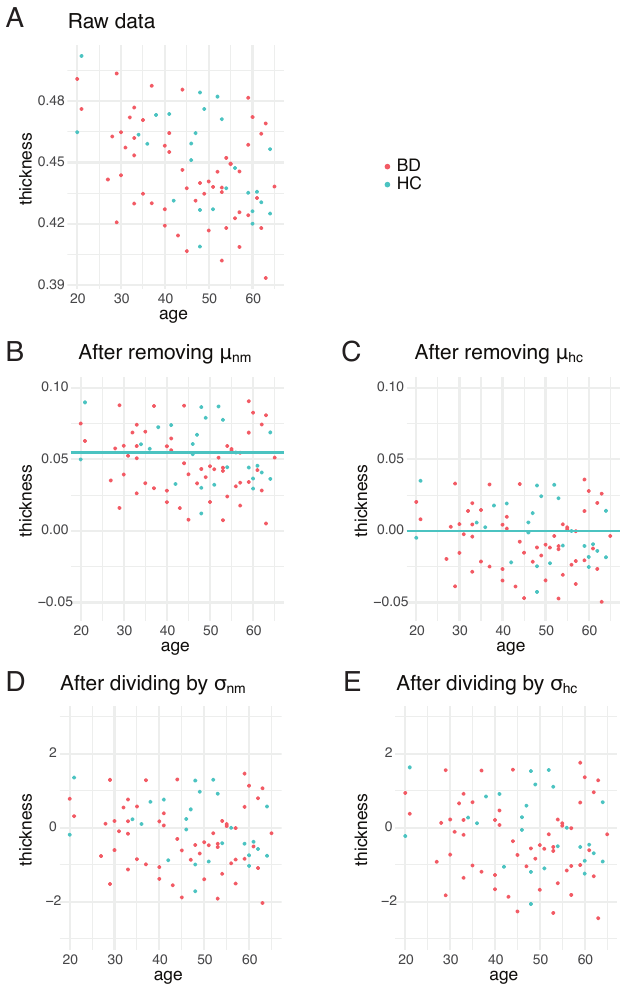}
  \caption{\textbf{Application of normative model to unseen data.} Plots show cortical thickness for one example region. \textbf{A}~Raw data from new scanning site. \textbf{B}~Data after removing the mean $\mu_{nm}$ estimated from the normative model (step 2.). The blue line indicates the healthy control mean $\mu_hc$. \textbf{C}~Data after removing the site-specific mean $\mu_{hc}$ (step 4.). \textbf{D}~Data after dividing by the standard deviation $\sigma_{nm}$ estimated from the normative model (step 5.). \textbf{E}~Data after dividing by the site-specific standard deviation $\sigma_{hc}$ (step 7.).}
  \label{apply}
\end{figure}

To apply the model to new datasets, we fitted it to the new data assuming it was from one of the scanning sites used for training, to predict model parameters and calculate residuals. This removed site and sex effects from the data. Site-specific mean and variance were then estimated as the mean and variance of the residuals of the new site's healthy controls. The exact steps of this algorithm are described in table \ref{apply}.

\begin{table}[bh!]
\centering
\begin{tabular}{ p{1cm}|p{5cm}|p{6cm} } 
 Step & Change & Explanation \\ 
 \hline
 \hline
 1 & Predict $\mu_{nm}, \sigma_{nm}, \nu_{nm}, \tau_{nm}$ for each subject from normative model. & Predict distribution parameters for each new subject based on their age and sex, but with their site set to one of the normative sites (site A). \\ 
 \hline
 2 & $y_1 = y - \mu_{nm}$ & Calculate residuals for each subject relative to site A by subtracting the predicted $\mu$, which removes age and sex effects and centres the new site around the mean of site A. \\ 
 \hline
 3 & Estimate $\mu_{hc} = $ mean($y_1$) from healthy controls. & Estimate the mean of the residuals from the healthy controls. This is the site-specific offset of the new site relative to site A. \\ 
 \hline
 4 & $y_2 = y_1 - \mu_{hc}$ & Calculate residuals for each subject relative to their site mean. \\ 
 \hline
 5 & $y_3 = y_2/\sigma_{nm}$ & Divide by the standard deviation predicted by the normative model, to z-score subjects relative to the variance in site A. \\ 
 \hline
 6 & Estimate $\sigma_{hc} = $ std($y_3$) from healthy controls \newline \textbf{or} \newline $\sigma_{hc} = $ mean($\sigma_{sites}$) & If the new site has more than 30 controls, estimate the standard deviation from the controls relative to site A. \newline \textbf{or} \newline If there are fewer than 30 controls, compute the mean of the site-specific standard deviations across normative sites (obtained as their $\sigma$ coefficients) as an estimate standard deviation instead. \\ 
 \hline
 7 & $zscore = y_3/\sigma_{hc}$ & The data is z-scored by dividing by the site-specific standard deviation.
\end{tabular}
\caption{Steps to apply the normative model to new data from an unseen scanning site. Here, $y$ is the measured data of one subject in a single cortical region in one metric (e.g. thickness). $y_n$ refers to the data after successive steps of manipulation. $\mu_{nm}, \sigma_{nm}, \nu_{nm}, \tau_{nm}$ are the mean, standard deviation, skew and kurtosis predicted for one subject by the normative model. $\mu_{hc}$ and $\sigma_{hc}$ are the site-specific mean and standard deviation relative to a normative site A.}
\end{table}

Here, steps~5-7 are similar to steps~2-4, but remove the standard deviation from the data rather than the mean. Whilst the mean of a dataset can be estimated reasonably reliably even from small samples, the estimation of standard deviation requires a larger sample. For this reason, we use even small ($>$10 subjects) healthy control data to estimate the new site's mean (step~3), but require at least 30~subjects for the estimation of the new site's standard deviation (step~6).

\begin{figure}[t!]
\centering
  \includegraphics[scale=1]{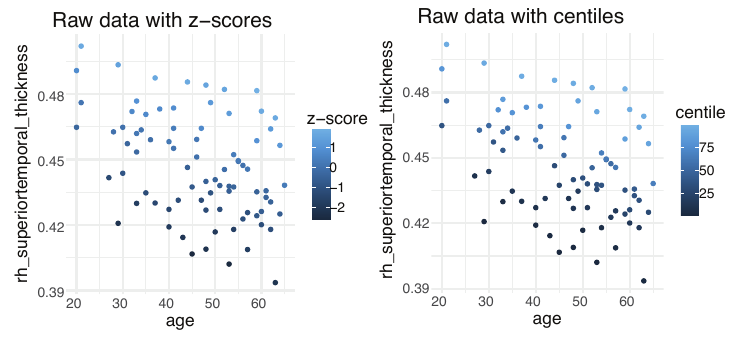}
  \caption{\textbf{Z-scores and centiles of unseen data.}}
  \label{z_c}
\end{figure}

\bigskip
To compute centiles, we estimate $\mu'_{hc} = $ mean($y_3$) again from the healthy controls. We then use that site mean, the site standard deviation ($\sigma_{hc}$), and the skew and kurtosis from the normative model ($\nu_{nm}$ and $\tau_{nm}$) to get quantiles for each subject and calculate their centiles (Figure \ref{z_c}).

\clearpage

\clearpage
\section{Assessing model stability through subsampling} \label{supplStable}

\begin{figure}[h!]
\centering
  \includegraphics{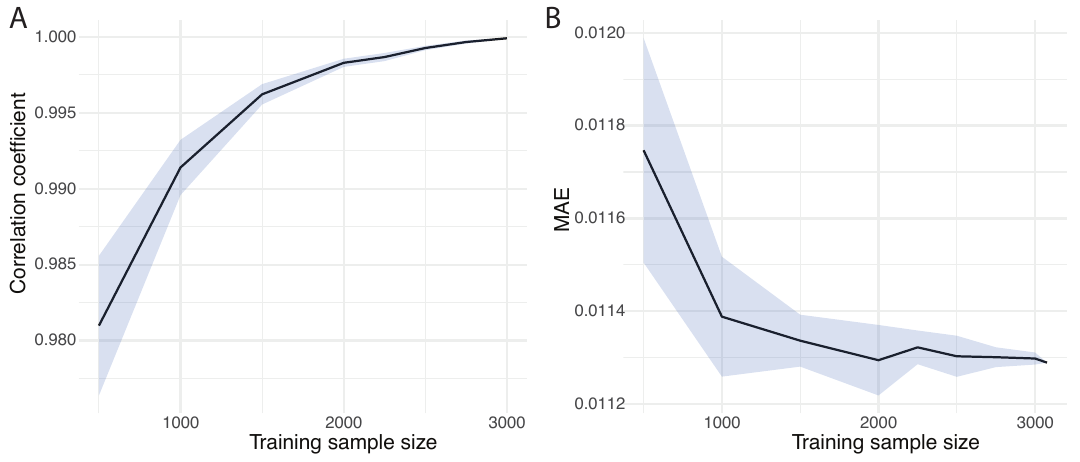}
  \caption{\textbf{Model stability trained on subsamples of full data.} Normative model of right hemisphere cortical thickness (log transformed) retrained on subsamples of the normative data set. 10 samples were repeated at each sample size. 200 subject randomly sampled across sites were held out to assess model similarity and fit across samples. \textbf{A)} Pearson correlation coefficient between residuals of 200 held-out subjects using subsampled model and full model (n=3076). Grey shading indicating 95\% confidence interval. \textbf{B)} Mean absolute error (MAE) of predictions of 200 held-out subject. Grey shading indicates 95\% confidence interval.}
  \label{subsample_CC_MAE}
\end{figure}

To assess the stability of our normative models, we tested correlations between models using a subset of the full data set and the convergence of errors on a random sample of held-out subjects. Figure \ref{subsample_CC_MAE}~A shows that models trained on sampled subsets of at least 2000 subjects have a high correlation ($>$99.5\%) with the model trained on the full dataset in terms of test set residuals, meaning predictions from the subset-trained model closely match those from the model using all available data. Figure \ref{subsample_CC_MAE}~B shows that the mean absolute error of predictions on the test set converges when using around 2000 subjects for training, indicating that model performance has little improvement with larger sample sizes. Together, these findings confirm that our model, trained on the full sample (n=3276), is stable.

\clearpage
\section{Validation using mesial temporal lobe epilepsy cohort}

\subsection{Comparing group-level cortical thickness abnormalities in mesial temporal lobe epilepsy with ENIGMA findings} \label{supplENIGMAcorr}

\begin{figure}[h!]
\centering
  \includegraphics[scale=.8]{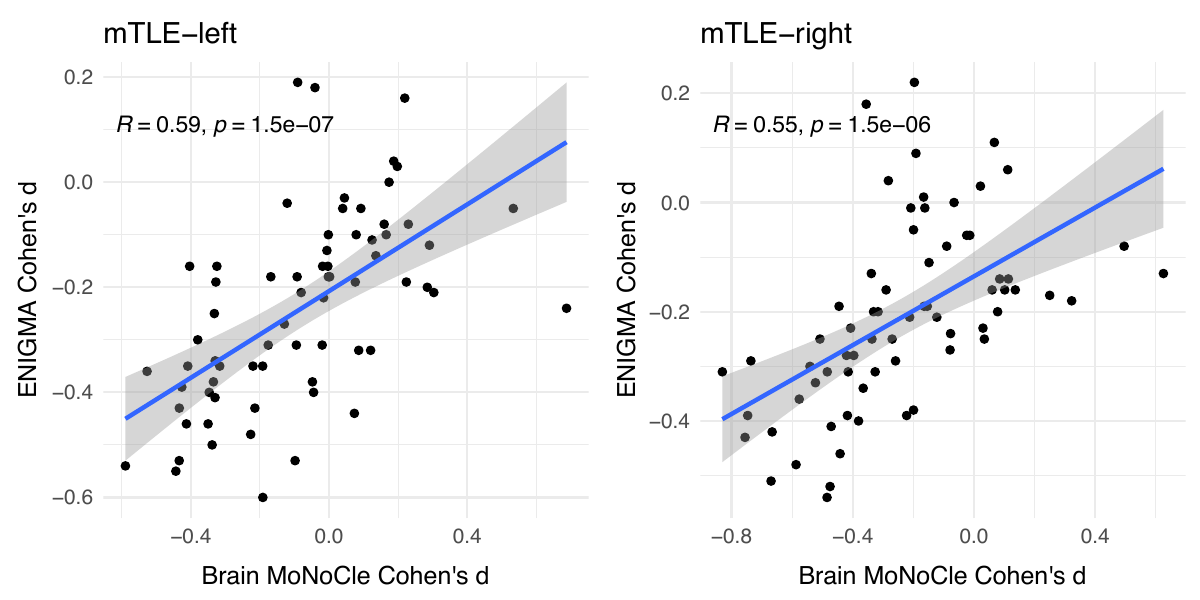}
  \caption{\textbf{Correlations of cortical thickness abnormalities of each brain region produced by Brain MoNoCle using the IDEAS dataset with equivalent effect sizes reported in the ENIGMA study.} Abnormalities were calculated as effect sizes (Cohen's d) when comparing each mTLE group to healthy controls. Left panel=mTLE left-lateralised, right panel=mTLE right-lateralised.}
  \label{enigma_corr}
\end{figure}

We correlated the effect size of the group difference (mTLE versus healthy controls) of cortical thickness of each brain region with the equivalent effect sizes reported in an ENIGMA study \parencite{Whelan2018}. This was done for mTLE-left and mTLE-right groups separately. Effect sizes produced by Brain MoNoCle in our sample show agreement (correlation larger than 0.5) with those reported previously, as shown in Figure \ref{enigma_corr}. Given that the ENIGMA sample of TLE was much larger, combining data from many more site, and using different data harmonisation methods, we believe that this level of agreement in the resultant effect sizes is remarkable.

\clearpage
\subsection{Prediction of TLE lateralisation} \label{supplLat}

\begin{figure}[h!]
\centering
  \includegraphics[scale=0.9]{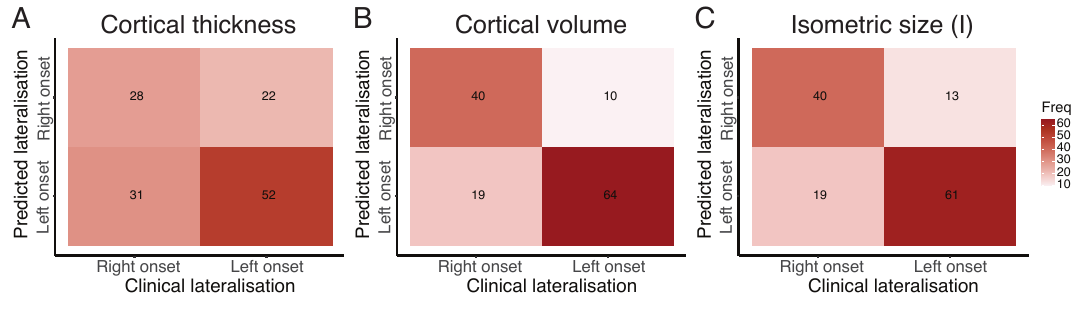}
  \caption{\textbf{Confusion matrices of predicted TLE lateralisation.} Performance of lateralisation prediction based on the sign of the difference of left and right hemisphere z-score. Confusion matrices are shown for metrics average cortical thickness (\textbf{A}), cortical volume (\textbf{B}), and isomectric size I (\textbf{C}).}
  \label{confMatr}
\end{figure}

We validated our normative model by assessing the agreement of a prediction of lateral hemisphere using abnormalities calculated from our normative model with clinical lateralisation in individuals with TLE. For three metrics (cortical thickness, cortical volume, and isometric size (I)), we predicted individuals' side of seizure onset to be the right hemisphere if the difference of left and right hemisphere z-scores (see Figure \ref{lateralisation}) was positive, and left if it was negative. Figure \ref{confMatr} shows a confusion matrix for each metric. Predictions using cortical volume performed best, achieving an accuracy 0.782.

\clearpage
\section{Comparison of Brain MoNoCle and CentileBrain} \label{supplCentBrain}
\begin{figure}[h!]
\centering
  \includegraphics[scale=0.5]{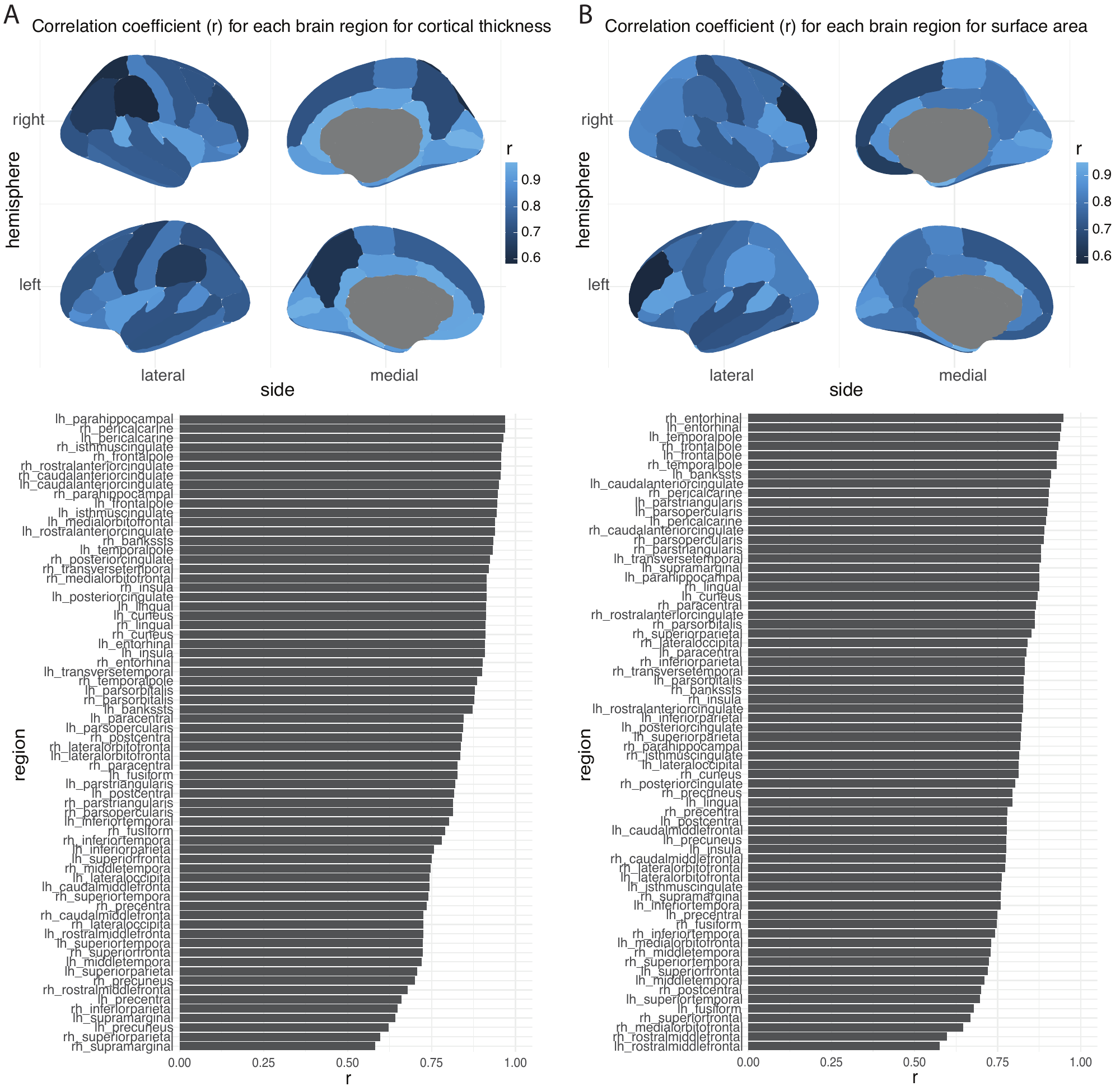}
  \caption{\textbf{Correlations of z-scores derived from Brain MoNoCle and CentileBrain.} Correlation coefficients are presented for each brain region for cortical thickness (\textbf{A}) and surface area (\textbf{B}).}
  \label{correlation_CB_NM}
\end{figure}

We ran data for healthy controls from the IDEAS dataset (n=99) through another normative modelling app, CentileBrain. We correlated the z-scores produced by CentileBrain and Brain MoNoCle to compare the output from each app. Figure \ref{correlation_CB_NM} illustrates the results of the correlations, which show good agreement between the two apps for most brain regions. 

\clearpage
\section{Additional Acknowledgements\label{supplaackn}}
Cambridge Centre for Ageing and Neuroscience (CamCAN). CamCAN funding was provided by the UK Biotechnology and Biological Sciences Research Council (grant number BB/H008217/1), together with support from the UK Medical Research Council and the University of Cambridge, UK.

OASIS-3 provided data: Longitudinal Multimodal Neuroimaging: Principal Investigators: T. Benzinger, D. Marcus, J. Morris; NIH P30 AG066444, P50 AG00561, P30 NS09857781, P01 AG026276, P01 AG003991, R01 AG043434, UL1 TR000448, R01 EB009352. AV-45 doses were provided by Avid Radiopharmaceuticals, a wholly-owned subsidiary of Eli Lilly.
HCP data were provided in part by the Human Connectome Project, WU-Minn Consortium (Principal Investigators: David Van Essen and Kamil Ugurbil; 1U54MH091657) funded by the 16 NIH Institutes and Centers that support the NIH Blueprint for Neuroscience Research; and by the McDonnell Center for Systems Neuroscience at Washington University. 

The Oregon ADHD1000 raw data is publicly available on the NIMH Data Archive (NDA) under \#1938.

We thank the NIMH Office of the Clinical Director, the outpatient behavioral health clinic and NMR center for providing support for the data collection. This work utilized the computational resources of the NIH HPC Biowulf cluster \url{http://hpc.nih.gov}.

We thank Anna Beres, Koryna Lewandowska, Monika Ostrogorska, Barbara Sikora-Wachowicz, Aleksandra Zyrkowska, Justyna Janik, Kamil Cepuch, and Piotr Faba for assistance with participant recruitment and data collection of chronotype dataset. They are funded by Polish National Science Centre (NCN) Grant 2013/08/M/HS6/00042 and Grant 2013/08/W/NZ3/00700.

Greene data collection was supported by National Institutes of Health Grants K01MH104592 (DJG), K23NS088590, UL1TR000448 (NUFD), R01MH096773, R00MH091238, UL1TR000128, MH110766, U01DA041148 (DAF), P30NS098577 (to the Neuroimaging Informatics and Analysis Center), and U54HD087011 (to the Eunice Kennedy Shriver National Institute Of Child Health \& Human Development of the National Institutes of Health to the Intellectual and Developmental Disabilities Research Center at Washington University; BLS), the Tourette Association of America (DJG), the McDonnell Center for Systems Neuroscience, the Mallinckrodt Institute of Radiology (DJG, NUFD), the DeStefano Family Foundation (DAF), the Jacobs Foundation, and the Child Neurology Foundation (NUFD).

% \section{Effects of mTLE in all metrics\label{mTLE_effect}}

% \begin{figure}
% 	\centering
% 	\includegraphics[width=\textwidth]{tle_all_metrics.png}
% 	\caption{\textbf{Heatmap illustrating the average abnormalities (z-scores) for each metric in mesial TLE groups.} CT=cortical thickness as exported by FreeSurfer.
% 	}
% 	\label{tle_all_metics}
% \end{figure}

%Left mTLE group showed abnormalities in both hemispheres (z-score smaller than -3), whereas right mTLE showed abnormalities in both hemispheres. In both groups, largest effects in the ipsilateral hemisphere were seen in isometric size I.

\newpage

\end{document}